\documentclass[aps,prl,twocolumn,showpacs,nobibnotes]{revtex4}

\usepackage{graphics,graphicx,amsfonts,amsmath,amsbsy,amssymb,color}
\usepackage{bm}
\usepackage{subfig}

\def \bfr {{\bf r}}

\def \bfi {{\bf i}}
\def \bfj {{\bf j}}
\def \bfk {{\bf k}}

\def \bfk {{{\bf k}}}

\def \bfq {{\bf q}}

\def \bfx {{\bf x}}

\def \beq {\begin{eqnarray}}
\def \eeq {\end{eqnarray}}

\def \Schrodinger {{Schr\"{o}dinger }}

\def \iFCIQMC {{\mbox{\emph{i}-FCIQMC }}}
\def \iFCIQMCbracket {{\mbox{\emph{i}-FCIQMC}}}
\def \infinity {{\infty}}

\def \bfj {{\bf j}}
\def \bfi {{\bf i}}

\def \ket {{\rangle}}
\def \bra {{\langle}}
\newcommand{\Hamil}{\hat{H}}

\newcommand{\refeq}[1]{{Eq.~\ref{#1}}}
\newcommand{\reffig}[1]{{Fig. \ref{#1}}}
\newcommand{\reftab}[1]{{Table \ref{#1}}}
\newcommand{\half}{\frac{1}{2}}

\begin{document}
\title{A Full Configuration Interaction Perspective on the Homogeneous Electron Gas}
\author{James~J.~Shepherd}
\email{js615@cam.ac.uk}
\affiliation{University of Cambridge, Chemistry Department, Lensfield Road, Cambridge CB2 1EW, U. K.}
\author{George~Booth}
\affiliation{University of Cambridge, Chemistry Department, Lensfield Road, Cambridge CB2 1EW, U. K.}
\author{Andreas~Gr\"{u}neis}
\affiliation{University of Cambridge, Chemistry Department, Lensfield Road, Cambridge CB2 1EW, U. K.}
\author{Ali~Alavi}
\email{asa10@cam.ac.uk}
\affiliation{University of Cambridge, Chemistry Department, Lensfield Road, Cambridge CB2 1EW, U. K.}
\begin{abstract}
Highly accurate results for the homogeneous electron gas (HEG) have only been achieved to date within a diffusion Monte Carlo (DMC) framework. Here, we introduce a newly developed stochastic technique, Full Configuration Interaction Quantum Monte Carlo (FCIQMC), which samples the exact wavefunction expanded in plane wave Slater determinants. Despite the introduction of a basis set incompleteness error, we obtain a finite-basis energy which is significantly, and variationally lower than any previously published work for the 54-electron HEG at $r_s$ = 0.5 a.u., in a Hilbert space of $10^{108}$ Slater determinants. At this value of $r_s$, as well as of 1.0 a.u., we remove the remaining basis set incompleteness error by extrapolation, yielding results comparable or better than state-of-the-art DMC backflow energies. In doing so, we demonstrate that it is possible to yield highly accurate results with the FCIQMC method in sizable periodic systems.
\end{abstract}
\date{\today}
\maketitle

The homogeneous electron gas (HEG), described by a simple model Hamiltonian, encapsulates many of the difficulties with modern electronic structure theory. To date the only truly successful \emph{ab initio} methods to yield accurate ground state energies at a range of densities have been quantum Monte Carlo techniques, in particular diffusion Monte Carlo (DMC) \cite{CeperleyAlder1980, Holzman1993, OrtizBallone1994,Kwon1998,Gurtubay2010,Rios}. The most famous of these was the results of Ceperley and Alder from which the LDA functionals of Density Functional Theory were parameterised \cite{PerdewZungerLDA,CeperleyAlder1980}. Diffusion Monte Carlo would be an exact technique but for the fixed-node approximation, which requires the nodes in the wavefunction due to fermionic exchange to be specified in advance by some trial wavefunction. In general, the fixed-node approximation lacks a method of being systematically improved to find the exact result. Attempts to go beyond the fixed-node approximation have been met with some success, however complete elimination of this error has not been achieved\cite{CeperleyAlder1980,Rios,Kwon1998,Holzman1993}. In particular, the release node (RN) method is practical only in systems for which the Bosonic ground state is close in energy to the Fermionic one. In the HEG this is only true at low density. At high densities, the RN-DMC is unstable, and fixed-node DMC with backflow corrections is the most viable option. This leaves open the question of the magnitude of the remaining fixed-node error.

Full configuration interaction (FCI) aims to find the wavefunction expressed as a linear combination of Slater determinants, formed from rearrangements of $N$ electrons in an underlying one-electron basis of $M$ spatial orbitals\cite{Olsen1988, Knowles1984}. This is equivalent to an exact diagonalization of this space. Since such a basis set of Slater determinants scales as $\binom{M}{N/2}^2$, benchmarks from FCI are extremely difficult to produce. There has been surprisingly little work undertaken with polynomially-scaling high-accuracy quantum chemical techniques, even though it has recently been shown that finite systems ranging from as few as 54 electrons can begin to capture the physics of the 3D HEG accurately\cite{Drummond2008,Holzmann2011}. In part this might be due to the required size of the one-electron basis and that, on approaching the thermodynamic limit for metals, many approximate methods find divergent energies\cite{Onsager}. In contrast, truncated configuration interaction will tend towards zero correlation energy.

We present the application of a new method, FCI quantum Monte Carlo\cite{FCIQMCPAPER1,FCIQMCPAPER2,EApaper,c2paper}, which stochastically samples the exact wavefunction providing the accuracy of exact diagonalization at a greatly reduced computational cost, to the high-density 54-electron HEG at $r_s$=0.5 and 1.0~a.u. This is the regime in which backflow corrections to FN-DMC are the largest\cite{Kwon1998,Rios}.

\emph{The Model.-} We seek to find the ground-state wavefunction and total energy of the $N$-electron HEG simulation-cell Hamiltonian:
\begin{equation}
\Hamil=\sum_\alpha -\half \nabla_\alpha^2 + \sum_{\alpha\neq \beta} \half \hat{v}_{\alpha\beta} + \half N v_\text{M}
\label{sim_cell_H}
\end{equation}
where the two-electron operator $\hat{v}_{\alpha\beta}$ is the Ewald interaction,
\begin{equation}
\hat{v}_{\alpha\beta}= \frac{1}{\Omega} \sum_\bfq v_\bfq e^{i \bfq \cdot \left( \bfr_{\alpha} - \bfr_\beta \right)} \quad ; \quad
v_\bfq = \left\{
\begin{array}{ll}
\frac{4\pi}{\bfq^2}, & \bfq\neq\bf{0} \\
0, & \mbox{\bfq=\bf{0}}
\end{array}
\right. 
\end{equation}
$v_\text{M}$ is the Madelung term, which represents contributions to the one-particle energy from interactions between a point charge and its own images and a neutralising background\cite{Ewald,Fraser1996,Drummond2008}, and $\Omega$ is the real-space unit cell volume. 

We use an expansion of the wavefunction in a Slater determinant basis, 
\begin{equation}
\Psi = \sum_\bfi C_\bfi | D_\bfi \ket,
\label{SDexp}
\end{equation}
where each determinant is a normalised, antisymmetrized product of plane waves,
\begin{equation}
D_\bfi = \mathcal{A} \left[ \psi_i(\bfx_1) \psi_j(\bfx_2) ... \psi_k(\bfx_N) \right]
\end{equation}
\begin{equation}
\psi_{j}  (\bfx) \equiv \psi_{j}  (\bfr , \sigma) =\sqrt{\frac{1}{\Omega}}~e^{i \bfk_j \cdot \bfr} ~\delta_{\sigma_j,\sigma}.
\end{equation}
The $\bfi$ index, which uniquely labels each determinant, is its normal-ordered string\cite{Kutzelnigg1997}.
The wavevectors $\bfk$ are chosen to correspond to the reciprocal lattice vectors of a real-space cubic cell of length $L$, 
\begin{equation}
\bfk=\frac{2\pi}{L} \left(n,m,l\right),
\end{equation}
where $n$,$m$ and $l$ are integers.
The Hartree-Fock determinant is the determinant occupying $N$ plane waves with the lowest kinetic energy. The full basis set for our calculation is constructed of all Slater determinants that can be made from $M$ plane waves ($2M$ spin orbitals) forming a closed-shell of orbitals in k-space with a kinetic energy lower than an energy cutoff $\half\bfk_c^2$. Plane waves are convenient because taking a single cutoff parameter to infinity makes the one-electron basis set complete. Moreover, plane waves are natural orbitals for the electron gas, implying that a FCI expansion is rapidly convergent in this basis\cite{Davidson}.

The determinant expansion given in \refeq{SDexp} can be inserted into the imaginary-time \Schrodinger equation, yielding a set of coupled equations for the determinant coefficients
\begin{equation}
-\frac{dC_\bfi}{d\tau} = (H_{\bfi\bfi}-S)C_{\bfi} + \sum_{\bfj \neq \bfi} H_{\bfi\bfj}C_{\bfj}.
\label{FCI-eqs}
\end{equation} 
Setting $dC_\bfi / d\tau = 0$ and solving for $S$ by exact diagonalization yields the total energy for the problem in a given basis.

\begin{figure}
  \centering
  \subfloat[A typical \iFCIQMC run. At $\tau\simeq 3.8$ a.u., the shift $S$ was allowed to vary to keep the walker number at an average of 20 million. From this point, an average was taken of the total energy.]{\label{MAINFIG1a}\includegraphics[width=0.5\textwidth]{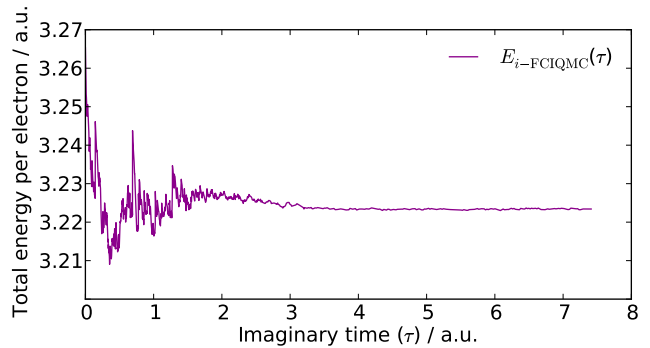}}
  
  \subfloat[\iFCIQMC calculations with $n_\text{add}$=3 are run at increasing $N_w$ values, with the aim that the limit $N_w \rightarrow \infty$ is found by the simulation at maximum walker number.]{\label{MAINFIG1b}\includegraphics[width=0.5\textwidth]{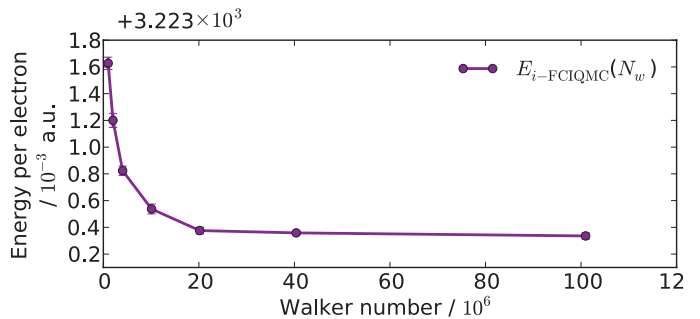}}
  
  \caption{Plots showing calculation of the \iFCIQMC energy for $N$ = 54, $2M$ = 682, $r_s$= 0.5 a.u. The result is reached in the limit of long time (iteration number) and large walker number $N_w$.}
  \label{typicalFCIMC}
\end{figure}

In a novel and recently developed quantum Monte Carlo, termed Full Configuration Interaction QMC (FCIQMC)\cite{FCIQMCPAPER1}, \refeq{FCI-eqs} is regarded as a set of master equations governing the dynamics of the evolution of the determinant coefficients in imaginary time, with elements of $\bf{H}$ being non-unitary transition rates. These dynamics are simulated by introducing a population of $N_w$ `walkers' distributed over the determinants, which are signed to represent the sign of the coefficients within the simulation, $C_\bfi \propto \bra N_\bfi \left(\tau\right) \ket$. The walker population is then allowed to evolve through discretized imaginary time-steps by spawning, death/cloning and annihilation events according to \refeq{FCI-eqs} until a steady-state is reached. The exact rules for this can be found in \cite{FCIQMCPAPER1}. 

The parameter $S$, termed the shift, is a population control parameter which can be updated self-consistently at equilibrium to oscillate around the total energy. However, throughout this work, the projected energy is used as a stochastic correlation energy estimator,
\begin{equation}
E_{\text{FCIQMC}} = \bra E(\tau) \ket=\sum_{\bfj} \bra D_\bfj | H | D_{\bf 0} \ket \frac{\bra N_{\bfj} (\tau)\ket}{\bra N_{\bf 0} (\tau)\ket},
\end{equation}
where $D_{\bf 0}$ is taken as the Hartree-Fock determinant and the sum $\bfj$ need only be taken over the $\mathcal{O} \left[ N^2 M \right]$ doubly excited determinants of $D_{\bf 0}$.

Typically the system is initially grown by setting $S$ to some positive value and allowing evolution from a single determinant to allow an unbiased evolution of the population. Only populations above a critical system-dependent size are able to converge to the FCI distribution, and this size scales linearly with the size of the Hilbert space\cite{FCIQMCPAPER1}. 

In order to alleviate this problem, an adaptation of this method has been developed, called  initiator-FCIQMC (\iFCIQMCbracket)\cite{FCIQMCPAPER2,EApaper,c2paper}. The determinant space is instantaneously divided into those determinants exceeding a population of $n_\text{add}$ walkers, termed initiator determinants, and those that do not. When considering a determinant whose current population is zero, the sum in the second term of \refeq{FCI-eqs}, the term describing net flux of walkers onto that determinant, is taken to be only over initiator determinants. This effectively introduces a survival of the fittest criterion for survival of newly spawned walkers. If a walker has been spawned from a determinant with an instantaneous population exceeding a parameter $n_\text{add}$, the child is allowed to survive. However, if the parent walker is a determinant with a population smaller or equal to $n_\text{add}$ then the child only survives if it has been spawned to a currently occupied determinant. This \iFCIQMC has been shown to dramatically accelerate the convergence of FCIQMC with respect to walker number. Note that in the large walker number limit, the \iFCIQMC tends to the FCIQMC algorithm, which itself converges rigourously to the FCI energy. Figure \ref{typicalFCIMC} illustrates an \iFCIQMC energy calculation in this way. Previous work has shown that the rapid convergence to this limit can be examined by finding correlation energies at increasing walker numbers (\reffig{MAINFIG1b})\cite{c2paper}.

\begin{figure}
\begin{center}
\includegraphics[width=0.5\textwidth]{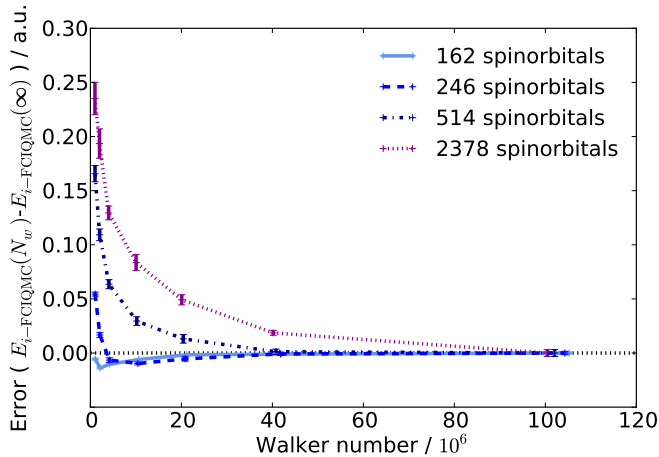}
 \end{center}
 
  \caption{Convergence with walker number up to approximately 100 million walkers (taken to be the infinite walker limit) is shown for a variety of basis sets for $N$ = 54, $r_s$= 1.0 a.u. Each line is labelled with the spinorbital number, $2M$ and was calculated with $n_\text{add}$=3. As the basis set size grows, so the size of space and the number of walkers required to sample the space accurately grows.}
  \label{INITFIG}
\end{figure}

As the basis set size grows, so the number of walkers required to recover the total energy to a given level of accuracy increases (\reffig{INITFIG}). Further results are taken to be converged with respect to this initiator error.

\emph{Basis set extrapolation.-} Although \iFCIQMC is able to produce exact results in a finite basis set, these are only upper-bound estimates to the true ground-state energy. This error in the energy, termed the basis set incompleteness error, is absent in DMC results which do not have a substantial dependence on a basis set\cite{Foulkes2001}.

\begin{figure*}
  \centering
  \subfloat[$r_s$=0.5]{\label{config1}\includegraphics[width=0.50\textwidth]{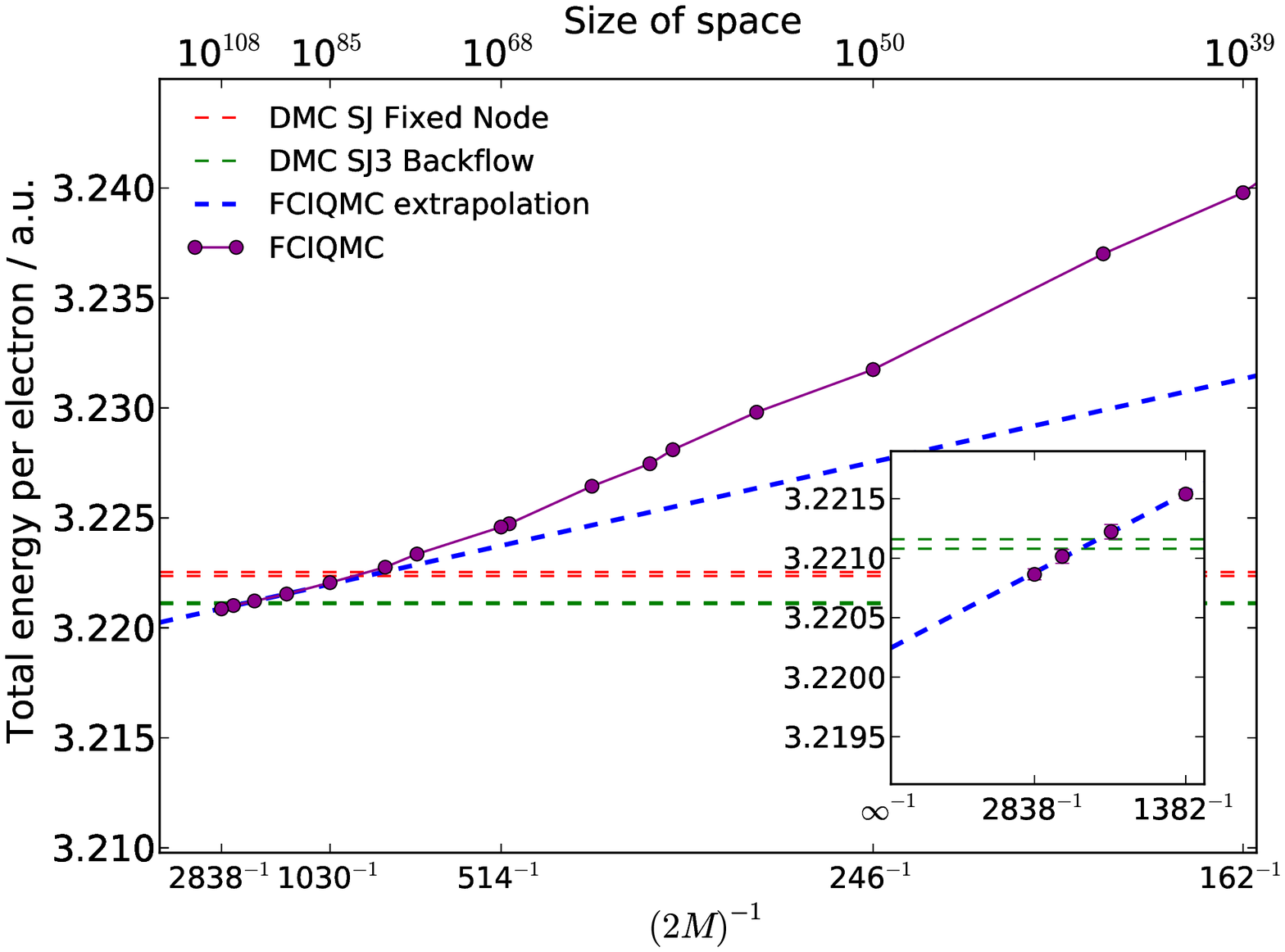}}
  \subfloat[$r_s$=1.0]{\label{config2}\includegraphics[width=0.50\textwidth]{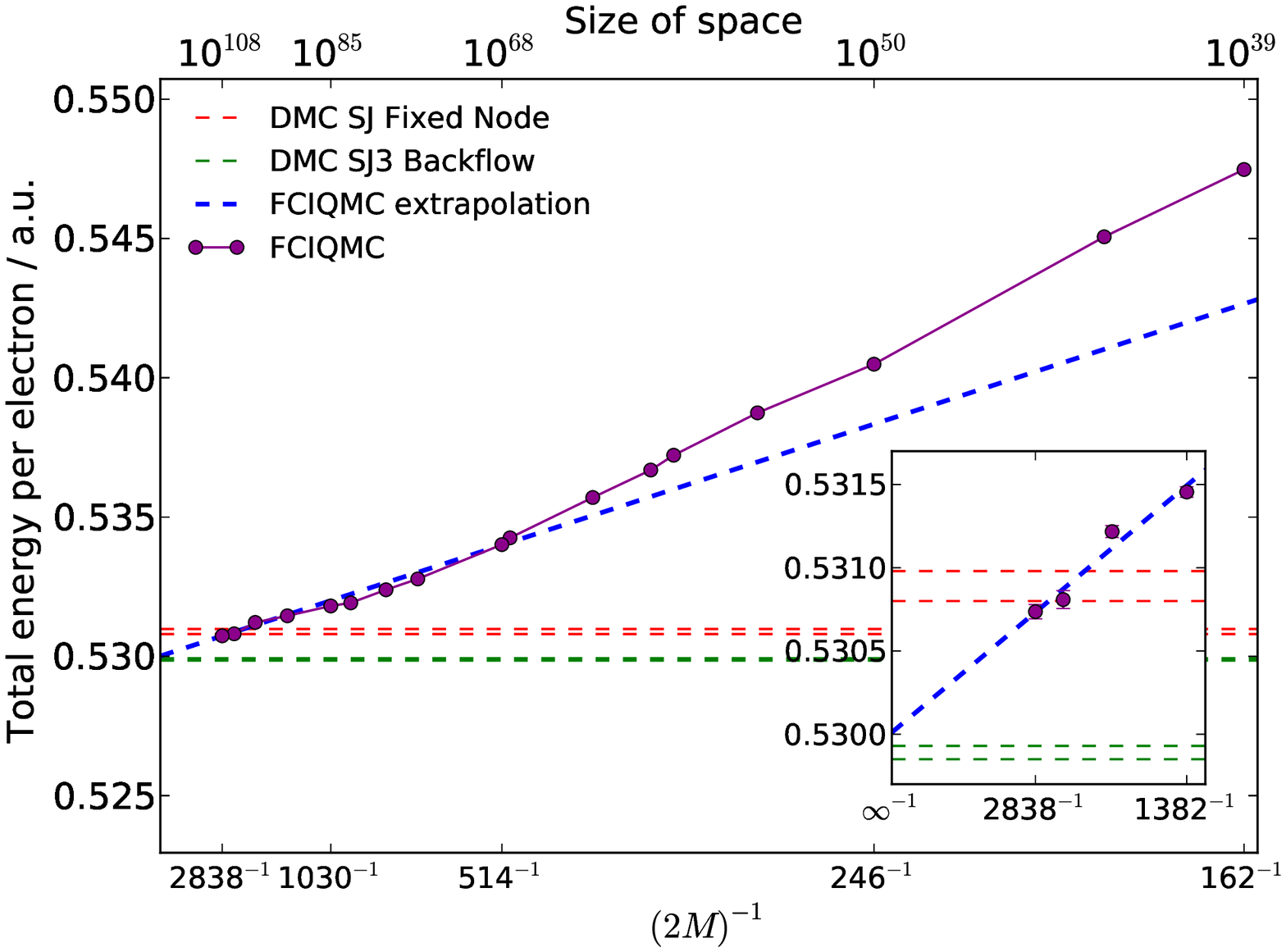}}
  
  \caption{\iFCIQMC total energies for a basis of $2M$ spin orbitals. Each basis set corresponds to a kinetic energy cutoff, with $2M=2838$ corresponding to 208~Ryd at $r_s$=0.5~a.u. and 52.1~Ryd at $r_s$=1.0~a.u. Each calculation used 40 million walkers for $r_s=0.5$~a.u. and 100 million walkers for $r_s=1.0$~a.u.. The blue dashed line is an extrapolation to $M\rightarrow\infinity$ based on the expected form $1/M$ using the data set with the largest number of walkers, shown with error bars in the inset. The DMC results, taken from R\'{\i}os \emph{et al.}\cite{Rios}, do not suffer from basis set error and are shown as two horizontal lines representing the mean plus and minus one standard deviation. Almost identical backflow results can be found for $r_s=1.0$~a.u. in a study by Kwon \emph{et al.}\cite{Kwon1998}.
  }
  \label{mainfigEXTRAP}
\end{figure*}

Extrapolation of the correlation energy to the complete basis set limit is performed regularly in molecular systems for which scaling laws have been investigated extensively\cite{Kutzelnigg}. In plane wave systems, a $1/M$ extrapolation is used for the basis set incompleteness error in methods employing the random phase approximation and second-order M\o ller-Plesset theory\cite{Harl2008, Gruneis2010}. We note that analytic expressions can be derived with these methods for the HEG that also show a $1/M$ relationship. Figure \ref{mainfigEXTRAP} illustrates that by using this fit at high basis set sizes, complete basis set exact diagonalization energies can be obtained that compare well with most recent high-accuracy DMC results\cite{Rios}. The unquantified initiator error is sub-m$\text{E}_\text{h}$, and is the largest source of error in these results but these are nonetheless thought to be upper-bound estimates of the exact energy, due to the observed variationality of the initiator error in large basis sets (see \reffig{INITFIG}).

\emph{Results and Conclusions.-} Results of \iFCIQMC calculations performed on the 54-electron gas, for basis sets containing between 162 and 2838 spinorbtials, are shown in \reffig{mainfigEXTRAP} and \reftab{onlytable}. In these calculations, Hilbert spaces ranging from $10^{39}$ to $10^{108}$ Slater determinants (\reffig{mainfigEXTRAP}) were sampled to produce high-accuracy energies for the HEG using approximately 300,000 core hours in total.

\begin{table}
\begin{tabular}{ c | c | c }
   $r_s$  & FCI $2M=2838$  & FCI $M=\infty$ \\
   / a.u. & / a.u. per electron & / a.u. per electron \\
  \hline      
  \hline  
  0.5 & 3.22086(2) & 3.2202(2) \\
  1.0 & 0.53073(4) & 0.5300(3) \\
  \hline  
\end{tabular}
  \caption{\iFCIQMC total energies for $2M$ spin orbitals. The error estimate for the finite-basis corresponds to stochastic error. The $M=\infty$ result is based on extrapolations shown in \reffig{mainfigEXTRAP}, from which the error estimate derives. }
  \label{onlytable}
\end{table}

For $r_s$=1.0~a.u., we obtain a variational finite-basis result that lies just below the fixed-node DMC result, with the extrapolated energy agreeing well with backflow DMC energies. At this $r_s$, the remaining fixed node error can be estimated by variance extrapolation of the backflow DMC results, and is thought to be sub-m$\text{E}_\text{h}$\cite{Kwon1998}. The results presented here, containing a similar order of magnitude of error, seem to substantiate this claim. To resolve the remaining fixed-node error in the backflow results, we would need to further reduce the leading source of error in the \iFCIQMC results, which is due to basis set incompleteness.

For $r_s$=0.5~a.u., we obtain a variational finite-basis result that lies below the backflow DMC result. However, our extrapolated energy falls significantly below the lowest DMC result found to date, suggesting residual fixed-node error in the backflow DMC energies is of the same order as the backflow corrections themselves. It has been commented that the wavefunctions for the 3D HEG change much more at higher densities than at lower densities under backflow transformations\cite{Kwon1998}. Our results suggest that in spite of this greater change, the backflow transformation still comparatively struggles at $r_s=0.5$~a.u.

The significance of these results extends beyond the sheer size of the many-electron basis which is being effectively sampled without error. The fact the results can be taken as exact within the designated basis set allows them to be used to benchmark other, more approximate methods in this system, as well as providing upper bounds to which the DMC method can optimise its nodal surface. Our method can also easily be extended to examine other properties of the HEG, in particular momentum distributions and Fermi liquid parameters, which is the focus of many current studies\cite{Holzman1993, OrtizBallone1994,Kwon1998,Gurtubay2010,Rios,Holzmann2011}. Beyond the HEG, the results also bode well for the treatment of more realistic solid-state systems, including metals, at unprecedented accuracy.

\begin{acknowledgements}
\emph{Acknowledgements.-} The authors gratefully acknowledge Neil Drummond for many useful conversations and to Mike Towler for providing us the CASINO code for comparison benchmark values at the early stages of this study\cite{Needs2010}. This work was supported by a grant from the Distributed European Infrastructure for Supercomputing Applications under their Extreme Computing Initiative.
\end{acknowledgements}


\end{document}